\documentclass[12pt]{article}
\usepackage{amsfonts}
\usepackage{amssymb}
\usepackage{epsfig}
\begin{document}
\newtheorem{theorem}{Theorem}
\newtheorem{lemma}{Lemma}
\newtheorem{definition}{Definition}
\newcounter{one}
\setcounter{one}{1}
\newcounter{six}
\setcounter{six}{6}
\newcounter{nineteen}
\setcounter{nineteen}{19}
\renewcommand{\Re}{\mathop{\mathrm{Re}}}
\renewcommand{\Im}{\mathop{\mathrm{Im}}}

\begin{center}
{\Large\bf Topological Phenomena in the Real Periodic Sine-Gordon
Theory}

 \vspace{0.3cm}

 {\large\bf P.G.Grinevich}\footnote{The author was
partially supported by the RFBR Grant No 01-01-00874-A. Numerical
simulations were partially performed on a computer, donated to
P.G.Grinevich by the
Humboldt Foundation.}\\
Landau Institute for Theoretical Physics \\
of the Russian Academy of Sciences, \\
Moscow 117940 Kosygin Street 2,\\
e-mail pgg@landau.ac.ru \\
\medskip
{\large\bf S.P.Novikov}\footnote{The author was partially supported by
 NSF Grant DMS 0072700} \\
IPST, University of Maryland-College Park, \\
MD 20742-2431 \\
and Landau Institute for Theoretical Physics, Moscow,\\
e-mail novikov@ipst.umd.edu
\end{center}

\section{Introduction. Algebro-Geometrical Solutions and the Reality
 Reductions. The topological charge.}

The algebro-geometrical (or finite-gap) solutions play a very
important role in the modern theory of soliton equations with
periodic and quasiperiodic boundary conditions. The main property
characterizing these solutions is the following: the wave function
of the auxiliary linear operator is meromorphic in spectral
parameter on the finite part of an algebraic Riemann surface. It
allows to apply the powerful tools of classical algebraic
geometry, and to write explicit representations for solutions and
wave functions in terms of Riemann $\Theta$-functions of several
variables. Usually the algebro-geometrical solutions are dense in
the space of the periodic ones, and if the nonlinearity tends to
zero, they degenerate to finite Fourier sums. Therefore their role
in the soliton theory is rather similar to the role of finite
Fourier series for the linear PDE's.

The algebro-geometrical solutions of solition equations were first
introduced in the paper \cite{NOV} in 1974 dedicated to the
integration of the Korteveg-de Vries (KdV) equation with the
periodic boundary conditions. Real non-singular periodic
algebro-geometrical KdV solutions can be characterised by the
following property: the number of gaps (forbidden bands or zones)
in the spectrum of the auxiliary Schr\"odinger operator acting in
the Hilbert Space $L_2(R)$ is finite. This characterization is not
valid for other systems (for example the spectrum of
non-selfadjoint auxiliary operator for the self-focusing Nonlinear
Schr\"odinger equation (NLS) has no gaps for any regular
potential), but neverthereless the algebro-geometrical solutions
are called finite-gap.

The complete theory of algebro-geometrical KdV solutions involving the
spectral theory of
periodic (quasiperiodic) finite-gap Schr\"odinger operators on the line and its
unification with algebraic geometry and analysis on the Riemann
surfaces, time dynamics and algebro-geometric hamiltonian aspects was
constructed in the works \cite{DN1,DN2,D1,IM1,L,McK-VM} (see in the
survey article \cite{DMN}) and in the works \cite{FMc,NV1}. Complex
finite-gap solutions of the NLS and Sine-Gordon (SG) equations were
constructed in
\cite{KK,IK}. Finite-gap solutions of the $2+1$ Kadomtsev-Petviashvili
(KP) eqation were consturcted in \cite{K1} on a basis of a purely algebraic
formulation of the finite-gap approach developed in this paper. The
algebro-geometric spectral theory of the stationary periodic 2D Schr\"odinger
operator was developed in the works \cite{DKN,NV}. In these
papers it was shown, that the proper analog of the finite-gap
constraint in the 2-dimensional case is the following: the restriction
of the Bloch variety to one energy level is algebraic, and the spectral
data is ``collected'' at this energy level only. Such potentials can be
treated as ``integrable'' at one energy. It means, that the wave functions for
this energy and arbitrary complex quaismomenta can be written
explicitly in terms of Riemann $\Theta$-functions, but the information
about the spectral properties at other energies is very restricted.

The algebro-geometrical approach allows to obtain explicit compact
formulas for solutions. But the $\Theta$-fucntions of several
variables are very complicated from the analytic point of view;
therefore the existence of such formulas does not lead sometimes
to the simple solution of the problem of selecting physically or
geometrically meaningful solutions among generic complex ones. For
example, it very easy to describe reality conditions for the KdV,
defocusing NLS or Sinh-Gordon equation
$$
\frac{\partial^2 u}{\partial \xi \partial\eta}=4\sinh u(\xi,\eta).
$$
(in all these cases the auxiliary linear operators are
self-adjoint). However, the reality reductions for the
self-focusing NLS or the Sine-Gordon equation found in \cite{Cher}
are rather nontrivial. It was shown in \cite{NV} that analogous
reductions are responsible for the absence of magnetic field it
the theory of the 2D Schr\"odinger operator. A good review of
reality reductions can be found if \cite{Nat}.

The Sine-Gordon equation
\begin{equation}
u_{tt}-u_{xx}+\sin u(x,t)=0
\label{SG-0.1}
\end{equation}

derived in the \Roman{nineteen} century geometry in the light-cone representation

\begin{equation}
u_{\xi\eta}=4\sin u, \ \ \ u=u(\xi,\eta),
\label{SG-0.2}
\end{equation}

where
\begin{equation}
x=2(\xi+\eta), \ \ t=2(\xi-\eta), \ \
\partial_\xi=2\partial_x+2\partial_t, \ \
\partial_\eta=2\partial_x-2\partial_t.
\label{SG-0.3}
\end{equation}
is one the most important systems of the soliton theory. One it's
distinguished properties is the folowing. Usually if we fix the
spectral curve (or equvalently all local conservation laws), then the
level set is a connected Jacoby torus. It is true for KdV, defocusing
and self-focusing NLS, the both real reductions of the KP equation.
But in the Sine-Gordon case the level set is a union of $2^m$ real Jacoby
tori, where $m$ denotes the number of pairs of negative branch points.
As a corollary, the result of averaging a generic functional depends
not only  on the spectral curve, but also on the component. This property
is also essential if we calculate the action-angle variables
(this problem was studied in \cite{DN,Nov}).

The averaged densities of conservation laws arose naturally as
hamiltonians of the Whitham equations. These equations describe
the evolution of parametrs of the asymptotic solutions in the form
of a slowly varying $N$-phase wavetrains. The Whitham equations
for the Sine-Gordon were derived in the paper \cite{EFMM}, but the
connectivity components of the level sets were not discussed. Let
us point out, that the Whitham equations are hyperbolic only in
the ``stable'' case, when all branch point are negative and the
number of components is always greater than 1.

It is possible to extract from the calculations of \cite{EFMM},
that for all classical local conservation laws  except the
topological charge the result of averaging is the same for all
components, but this property was never pointed out explicitly. In
Section~\ref{section-averaging} we give a simple explanation of
it.

{\bf Remark.} The number of real tori was calculated in \cite{Cher},
their characterization in terms of the complex Jacoby torus was obtained
in \cite{Dubr-Nat}, see also \cite{Erc-Forest}. An elementary
description of these tori was obtained by the authors in  \cite{GN,GN2},
see Section~\ref{section-SG} below.

Another interesting property of the Sine-Gordon equation is the
following. In the theory of constant negative curvature surfaces
$u(\xi,\eta)$ denotes the angle between asymptotic lines. In the theory
of Josephson jucntions $u$ has the meaning of a phase. In both
situations it is defined up to an additive constant $c=2\pi n$.
Therefore it is natural to call a solution $u(x,t)$ space-periodic
with the period $T$ if $\exp\{iu(x+T)\} =\exp\{iu(x)\}$. The integer
$n$ such, that $u(x+T)=u(x)+2\pi n$ is called {\bf Topological charge},
the real number $\bar n=n/T$ denotes the {\bf Density of topological
charge}. These quantities are the ``most stable'' conservation laws
surviving all real periodic nonintegrable perturbations. The density
of topological charge can be naturally extended to all nonsingular
quasiperiodic solutions using the formula
$$
\bar n=\lim\limits_{T\rightarrow\infty}[u(x+T)-u(x)]/2\pi T.
$$
One of the most natural problems is the following: {\bf How to
calculate the topological charge of real periodic SG solutions in
terms of the Spectral Data: the Riemann surface and the divisor?}
It turned out, that nobody succeeded to extract the answer from
the explicit description of the real components described in terms
of the Jacoby torus and $\Theta$-functional formulas, found in the
works \cite{Dubr-Nat,Erc-Forest}. An attempt has been made almost
20 years ago in the work \cite{DN} to solve this problem. For this
purpose the so-called ``Algebro-Topological'' approach was
developed in \cite{DN}, and an interesting formula has been
proposed. However, as it was pointed out in \cite{Nov}, the idea
of the proof presented in \cite{DN} works only if the spectral
curve is sufficiently close to a degenerate one. A complete
solution using a new development of the main idea of the work
\cite{DN} was obtained only recently by the authors in the papers
\cite{GN} (for the so-called stable curves) and \cite{GN2} for
generic ones. Our proof is based on a new effective description of
the real components in terms of divisors. We present a summary of
these results in Section~\ref{section-tc}.

{\bf Remark.} Formally we can define topological charge for any
non-singular complex solution. But if the solution has poles, this
quantity became ill-defined, and no natural non-singularity
condition in terms of the spectral data for complex potentials is
known. It explains why only real solutions are discussed.

\section{Finite-gap Sine-Gordon solutions.}
\label{section-SG}
The ``soliton-type'' solutions of Sine-Gordon were found already in the
\Roman{nineteen} century using substitution discovered by Bianchi and
S.Lie. Now this method is called ``Backlund transformation''.

In the early 1970's G.Lamb found out (see \cite{Lamb}), that the
the Sine-Gordon equation is analogous to KdV in the following
sence: the so-called ``inverse scattering method'' can be applied
to it. The modern approach was started in the work \cite{AKNS}.
The critical point of this approach is the following
zero-curvature representation for SG
\begin{equation}
\Psi_x=\frac14 (U+V)\Psi, \ \ \Psi_t=\frac14(U-V)\Psi,
\label{SG-1.1}
\end{equation}
where
\begin{eqnarray}
U=U(\lambda,x,t)=\left[\begin{array}{cc}
i(u_x+u_t) & 1 \\ -\lambda  & -i(u_x+u_t)
\end{array}\right], \\
V=V(\lambda,x,t)=\left[\begin{array}{cc}
0 & -\frac{1}{\lambda}e^{iu} \\ e^{-iu} & 0.
\end{array}\right].
\label{SG-1.2}
\end{eqnarray}

To construct generic complex finite-gap SG solutions assume, that we
have the following spectral data:
\begin{enumerate}
\item A nonsingular hyperelliptic Riemann surface
    $\Gamma$ $[\mu^2=R(\lambda)]$, where
      $R(\lambda)=\prod\limits_{k=0}^{2g}(\lambda-E_k)$, such that $E_0=0$
      and $E_i\neq E_j, i,j=0,1,\ldots,2g$.
      It has exactly $2g+2$ branch points $E_0=0$, $E_1$, \ldots,
      $E_{2g}$, $\infty$; the genus of $\Gamma$ is equal to $g$. A point
      $\gamma\in\Gamma$ is by definition a pair of complex numbers
      $\gamma=(\lambda,\mu)$ such, that $\mu^2=R(\lambda)$.

\item A divisor $D$ of degree $g$, i.e. set (or formal sum) of $g$ points
$D=\gamma_1+\ldots+\gamma_g$.
\end{enumerate}

 For generic data $\Gamma$, $D$ there exists an
unique two-component ``Baker-Akhiezer''
 vector-function $\Psi(\gamma,x,t)$ such that
\begin{enumerate}
\item For fixed $(x,t)$ the function $\Psi(\gamma,x,t)$ is meromorphic
    in the variable
      $\gamma\in\Gamma$ outside the points $0$, $\infty$ and has at
      most 1-st order poles at the divisor points $\gamma_k$, $k=1,\ldots,g$.
\item  $\Psi(\gamma,x,t)$ has essential singularities at the points
      $0$, $\infty$ with the following asymptotics:
\begin{equation}
\Psi(\gamma,x,t)=\left(\begin{array}{c}
1+o(1) \\ i\sqrt{\lambda}+O(1)
\end{array} \right)e^{i\sqrt{\lambda}\frac{x+t}{4}} \ \ \mbox{as}
\ \ \lambda\rightarrow\infty,
\label{SG-1.3}
\end{equation}
\begin{equation}
\Psi(\gamma,x,t)=\left(\begin{array}{c}
\phi_1(x,t)+o(1) \\ i \sqrt{\lambda} \phi_2(x,t)+O(\lambda)
\end{array} \right)e^{-\frac{i}{\sqrt{\lambda}}\frac{x-t}{4}} \ \ \mbox{as} \ \
\lambda\rightarrow 0,
\label{SG-1.4}
\end{equation}
where $\phi_1(x,t)$, $\phi_2(x,t)$ are some functions of
the variables $x,t$ .
\end{enumerate}

Denote the divisor of zeroes of the first component
$\psi_1$ by $D(x,t)=\sum_j\gamma_j(x,t)$. The vector-function
$\Psi(\gamma,x,t)$  satisfies to the zero-curvature equations
(\ref{SG-1.1}) with  potential given by the formula
\begin{equation}
u(x,t)=i\ln \frac{\phi_2(x,t)}{\phi_1(x,t)},
\label{SG-1.5}
\end{equation}
and the function $u(x,t)$ solves the SG equation (\ref{SG-0.2}).
In terms of the divisor of zeroes $D(x,t)$ the potenital $u(x,t)$ can be
easily written explicitly:
\begin{equation}
e^{iu(x,t)}=\frac{\prod\limits_{j=1}^{g}(-\lambda_j(x,t))}
{\sqrt{\prod\limits_{j=1}^{2g} E_j}}.
\label{SG-1.6}
\end{equation}
The $x$ and $t$ dynamics of the divisor $D(x,t)$ can be described it
terms of Dubrovin equations (see \cite{GN2}). From (\ref{SG-1.6}) it
follows, that the potential $u(x,t)$ is singular at the point $(x_0,t_0)$
if and only if one of the points $\gamma_k(x_0,t_0)$ coincides with $0$
or $\infty$. Collisions of 2 or more divisor points result in
singularities in the solutions of Dubrovin equations, but all symmetric
combinations of the divisor points including the potential $u(x,t)$
remain non-singular.

{\bf Remark.} The first component $\psi_1$ of the Baker-Akhiezer
vector-function $\Psi$ is a partial case of the general scalar two-point
Baker-Akhiezer functions satisfying to the second order linear
Schr\"odinger equation $L_1\psi_1=0$ invented in the work
\cite{DKN}. Here we have
$$L_1=\frac{\partial^2}{\partial \xi\partial\eta} +A(\xi,\eta)\frac{
\partial }{\partial \eta} +W(\xi,\eta),$$ $-A=\partial_\xi\log\phi_1$.
In this special case our surface $\Gamma$ is hyperelliptic,  the
selected points coincide with $0,\infty$ and corresponding local
parameters are chosen as $\sqrt{\lambda},1/\sqrt{\lambda}$. The
second function $\psi_2$ satisfies to similar equation
$L_2\psi_2=0$ where the operator $L_2$ is obtained from $L_1$ by
the  so-called Laplace Transformations  and vice versa. It is
known that cyclic Laplace Chains of the length 2 lead to the SG
equation in the general complex case (see the work \cite{NV2}).

\begin{lemma}
Assume, that the spectral data $(\Gamma,D)$,
$D=\gamma_1+\ldots,\gamma_g$ satisfy the following reality constraints:
\begin{enumerate}
\item  The set of branch points of the Riemann surface $\Gamma$ contains
       real points or  complex conjugate pairs of points only, and all real
       branch points are nonpositive. Without loss of generality
       we may assume, that the first $2m+1$ of them
       $E_0=0$, $E_1$, $E_2$,\ldots, $E_{2m}$ are
       real and $0>E_1>E_2>\ldots>E_{2m}$, and for $k>m$ we have
       $E_{2k}=\overline{E_{2k-1}}$, $\Im E_{2k}\ne 0$. Then the spectral curve
       $\Gamma$ is real and admits a ${\mathbb Z}_2\times{\mathbb Z}_2$-group
       of involutions generated by the standard holomorphic involution
       \begin{equation}
    \sigma:\Gamma\rightarrow\Gamma:\ \  \sigma(\lambda,\mu)=(\lambda,-\mu),
     \label{SG-1.7}
       \end{equation}
       transposing the sheets, and by the antiholomorphic involution
       \begin{equation}
    \tau:\Gamma\rightarrow\Gamma:\ \ \tau(\lambda,\mu)=(\bar\lambda,\bar\mu),
     \label{SG-1.8}
       \end{equation}
       (complex conjugation).
 \item There exists a  meromophic differential $\Omega$ with two simple
       poles at the points $0,\infty$ such, that the zeroes of $\Omega$
       are located at the points $D+\tau D$. We
       shall call such divisors {\bf admissible}. Without loss of
       generality we can normalize $\Omega$ to be real: $\tau\Omega=\overline\Omega$.

\end{enumerate}
Then the potential $u(x,t)$ is real and non-singular.
\end{lemma}

{\bf Remark} The constraints on the spectral curve we found in
\cite{IK}. The characterisation of admissible divisors and the proof of
non-singularity were obatined in the work \cite{Cher}.

It is easy to check (see, for example \cite{GN2})
\begin{lemma}
If $D$ is admissible divisor, then the divisor of zeroes
$D(x,t)=\gamma_1(x,t)+\ldots+\gamma_g(x,t)$ is admissible for all $x,t$
\end{lemma}

Assume, that we have a real meromorphic differential $\Omega$ with
exactly 2 simple poles at the points $0,\infty$. We can associate an
admissible divisor $D$ to it if and only if $2g$ zeroes of $\Omega$
can be can be presented as an union of 2 subsets $D$ and $D_1$ such,
that $D_1=\tau D$. It is possible if and only if all real roots of
$\Omega$ have even multiplicity.

\begin{definition}
Assume that the spectral curve $\Gamma$ satisfy the reality condinitions
formulated above. A real meromorphic differential $\Omega$ on $\Gamma$ with
exactly 2 simple poles at the points $0,\infty$ is called {\bf admissible}
if the multiplicity of all real roots is even.
\end{definition}

Without loss of generality we may assume that the residued of $\Omega$
at the points $0$ and $\infty$ are equal to $+1$ and $-1$
respectively. Then we have

\begin{equation}
\label{SG-1.9}
\Omega=\left(1-\frac{\lambda P_{g-1}(\lambda)}{R(\lambda)^{1/2}}\right)
\frac{d\lambda}{2\lambda},
\end{equation}
where $P_{g-1}(\lambda)$ is a polynomial of degree at most $g-1$
with real coefficients. Let us call the polynomial $P_{g-1}(\lambda)$
{\bf admissible} if $\Omega$ is an admissible differential.

Denote the map associating an admissible polynomial with an admissible
divisor by $\Pi:D\rightarrow P_{g-l}(\lambda)$. The
inverse map $\Pi^{-1}$ is multivalued. If $P_{g-1}(\lambda)$ is an
admissible polynomial, then the equation
\begin{equation}
\label{SG-1.10}
\left\{
\begin{array}{c}
\mu=\lambda P_{g-1}(\lambda) \\
R(\lambda)=\mu^2,
\end{array}\right.
\end{equation}
has $2g+1$
roots. One of them is the point $(0,0)$, other $2g$ roots form $g$
pairs. To define an admissible divisor we have to choose one point
from each pair: therefore we have at most $2^g$ possibilities
depending on the number of real points. In the generic case we
have no real roots, so the number is equal to $2^g$ in this case.

The description of connected components suggested by the authors in \cite{GN,GN2}
is based on the following simple observation:

Consider a pair of functions of the real variable $\lambda$:
\begin{equation}
f_{\pm}(\lambda)=\pm\frac{\sqrt{R(\lambda)}}{\lambda}.
\label{SG-1.11}
\end{equation}
We assume, that the functions $f_{\pm}(\lambda)$ are defined only
for such $\lambda\ne0$ that $R(\lambda)\ge0$, i.e. these functions
are real-valued. Let us draw the graphs of these functions
$y=f_+(\lambda)$, $y=f_-(\lambda)$, and  fill in the following
domains:
\begin{equation}
\begin{array}{l}
\lambda<0, y^2 < \frac{R(\lambda)}{\lambda^2}, \\
\lambda>0, y^2 > \frac{R(\lambda)}{\lambda^2}.
\end{array}
\label{SG-1.12}
\end{equation}

\vspace{1cm}

\begin{center}
\mbox{\epsfxsize=10cm \epsffile{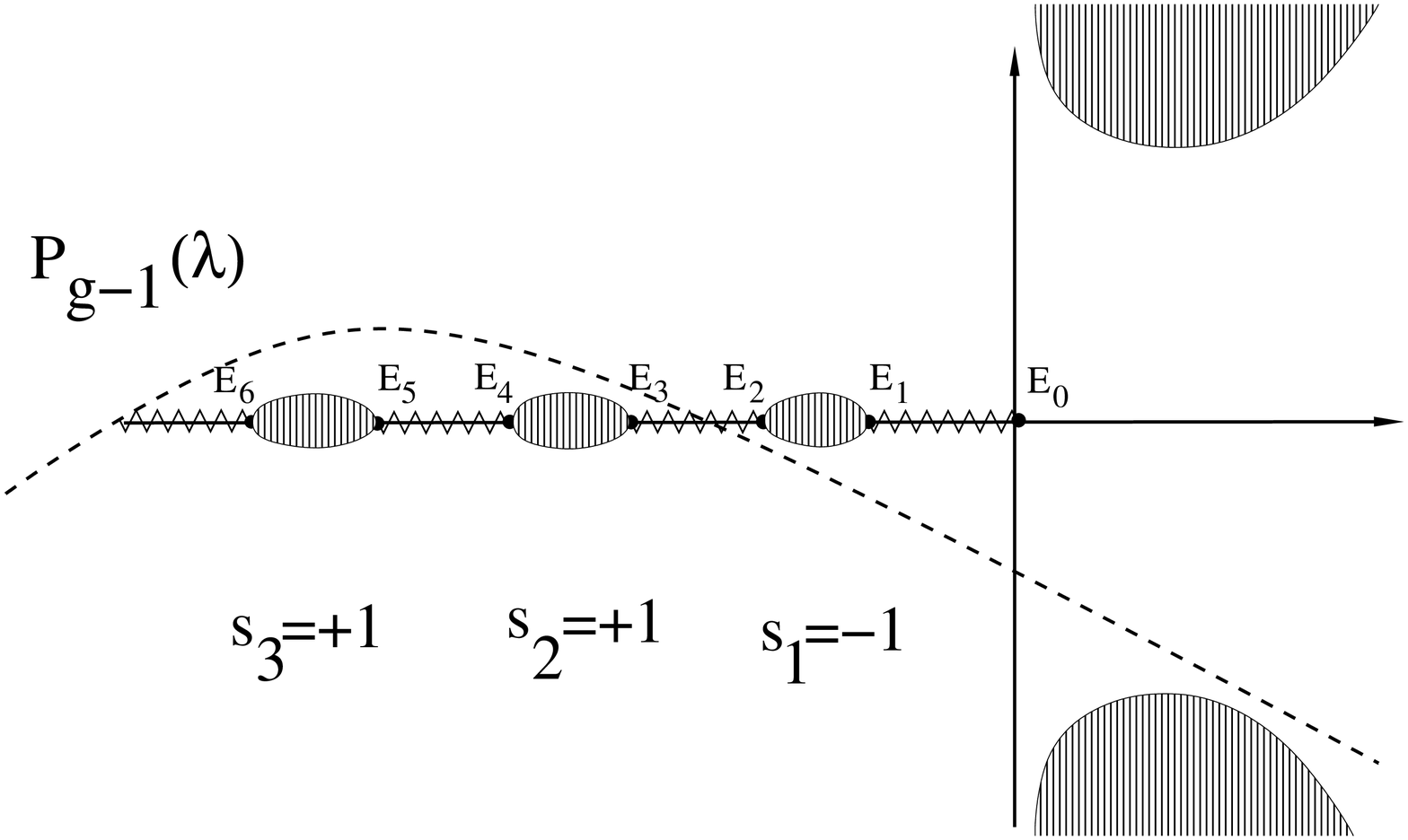}}

Fig 1.
\end{center}

\begin{lemma}
\label{SG-L1} The polynomial $P_{g-1}(\lambda)$ is admissible if
and only if the graph of $P_{g-1}(\lambda)$ does not cross the
black open domains (but it can touch their boundaries).
\end{lemma}

Now we can associate with each admissible polynomial a {\bf
Topological Type}, i.e. a collection of $m$ numbers $s_k=\pm1$,
$k=1,\ldots,m$ defined by the following rule:

$s_k=1$ if $P_{g-1}(\lambda) \ge f_+(\lambda)$ as
$E_{2k}\le\lambda\le E_{2k-1}$,

$s_k=-1$ if $P_{g-1}(\lambda) \le f_-(\lambda)$ as
$E_{2k}\le\lambda\le E_{2k-1}$.

Using  simple algebraic estimates (see \cite{GN2}) it is easy
to check:

\begin{lemma}
\label{SG-L1.1} Each topological type is presented by the convex
subset in the space of polynomials $P_{g-1}$. Each of these $2^m$
connected components is non-empty and depends continuously on the
branch points $E_1$,\ldots,$E_{2g}$.
\end{lemma}

The $x$ and $t$ dynamics of admissible divisors may be rather
non-trivial (a number of numeric experiments are discussed in
\cite{Erc-Forest}). In contrast with the self-ajoint case (KdV,
defocusing NLS, sinh-Gordon) the admissible position of a given
divisor point depends on the position of other ones (the admissibility
constraints are non-local). The trajectory of a
single divisor point may be not well-defined because divisor points  may
collide and a small variation of parameters near the collision point
results in a bifurcation of trajectories. If the potential $u(x,t)$ is
periodic with the period $T$, the corresponding divisor is generically
periodic only modulo permutations. But the following properties can be
easily proved using the characterization formulated above.

\begin{lemma}
\label{SG-L2} Projections of the points of admissible divisors to
the $\lambda$-plane could not lie in the segments $[E_1,E_0]$,
$[E_3,E_2]$,\ldots,$(-\infty,E_{2m}]$. Moreover, it is possible to
choose an open neighborhood of these segments such, that the
projections of these points could not lie in it.
\end{lemma}
\begin{lemma}
\label{SG-L2.1} Assume, that the projection of a point
$\gamma_s=(\lambda_s,\mu_s)$ of an admissible divisor to the
$\lambda$-plane lie in the segments $[E_{2k},E_{2k-1}]$; then $\mu_s<0$
if $s_k=1$ and $\mu_s>0$ if $s_k=-1$.
\end{lemma}

\section{The topological charge of real solutions}
\label{section-tc}

It is well-known (see, for example \cite{Dubr-Nat,Erc-Forest})
that after the Abel transform the $x$-dynamics of Sine-Gordon
corresponds to the motion along the straight line
\begin{equation}
\label{SG-2.1}
\vec X=\vec X_0+\vec U x,
\end{equation}
in the Jacoby torus, where $\vec U$ denotes the noramalized vector of
$b$-periods of the quasimomentum differential $dp$.
\begin{equation}
\label{SG-2.2}
U^k=-\frac{1}{2\pi}\oint\limits_{b_k}dp.
\end{equation}
Let us recall, that by definition $dp$ is a meromorphic differential
with zero $a$-periods and two second order poles at $0$, $\infty$ such, that
\begin{equation}
\label{SG-2.3}
dp=\left\{
\begin{array}{l}
\left(\frac{\displaystyle 1}{\displaystyle 8\sqrt{\lambda}}+
o\left(\frac{\displaystyle 1}{\displaystyle\lambda}\right)\right)d\lambda
\ \ \mbox{as} \ \ \lambda\rightarrow\infty,
\\ \\
\left(\frac{\displaystyle 1}{\displaystyle 8\lambda \sqrt{\lambda}}+
o\left(\frac{\displaystyle 1}{\displaystyle \lambda}\right)\right)d\lambda
\ \ \mbox{as} \ \ \lambda\rightarrow 0.
\end{array}
\right.
\end{equation}
The real part of the Jacoby torus is isomorphic to the factor
${\mathbb R}^n/{\mathbb Z}^n$, therefore we can apply the following
simple analytic lemma:

\begin{lemma}
\label{SG-L2.2}
 Let $u(\vec X)$, $X\in{\mathbb R}^n$ be a smooth function in ${\mathbb R}^n$
 such, that $\exp(iu(\vec X))$ is single-valued on the torus
 ${\mathbb R}^n/{\mathbb Z}^n$, i.e. $\exp(iu(\vec X+\vec N))=\exp(iu(\vec X))$
 for any integer vector $\vec N$. Denote by $u(x)$ restriction of
 $u(\vec X)$ to the straight line $\vec X=\vec X_0 +x\cdot\vec U$.
 Then the density of topological charge $\bar n=\lim\limits_{T\rightarrow\infty}
 [u(x+T)-u(x)]/2\pi T$ is well-defined; it does not depend on the
 point $\vec X_0$ and can be expressed by the following formula:
\begin{equation}
\bar n=\sum\limits_{k=1}^n n_k U^k,
\label{SG-2.4}
\end{equation}
where $\vec U=(U^1,U^2,\ldots,U^n)$, and $n_k$ are topological
charges along the basic cycles ${\mathcal A}_k$, $k=1,\ldots,n$:
\begin{equation}
\ \ \ \ \ \ \ \ \ \
u(X^1,X^2,\ldots,X^k+1,\ldots,X^n)-u(X^1,X^2,\ldots,X^k,\ldots,X^n)=2\pi n_k.
\label{SG-2.5}
\end{equation}
\end{lemma}

The main step is the calculation of the the charges along the basic
cycles ${\mathcal A}_k$. To obtain a convenient answer the choice of
canonical basis is very important. For {\bf stable} surfaces
(surfaces without complex branch points) the proper system of $a$-cycles
was constructed in \cite{DN}.

\begin{center}
\mbox{\epsfxsize=10cm \epsffile{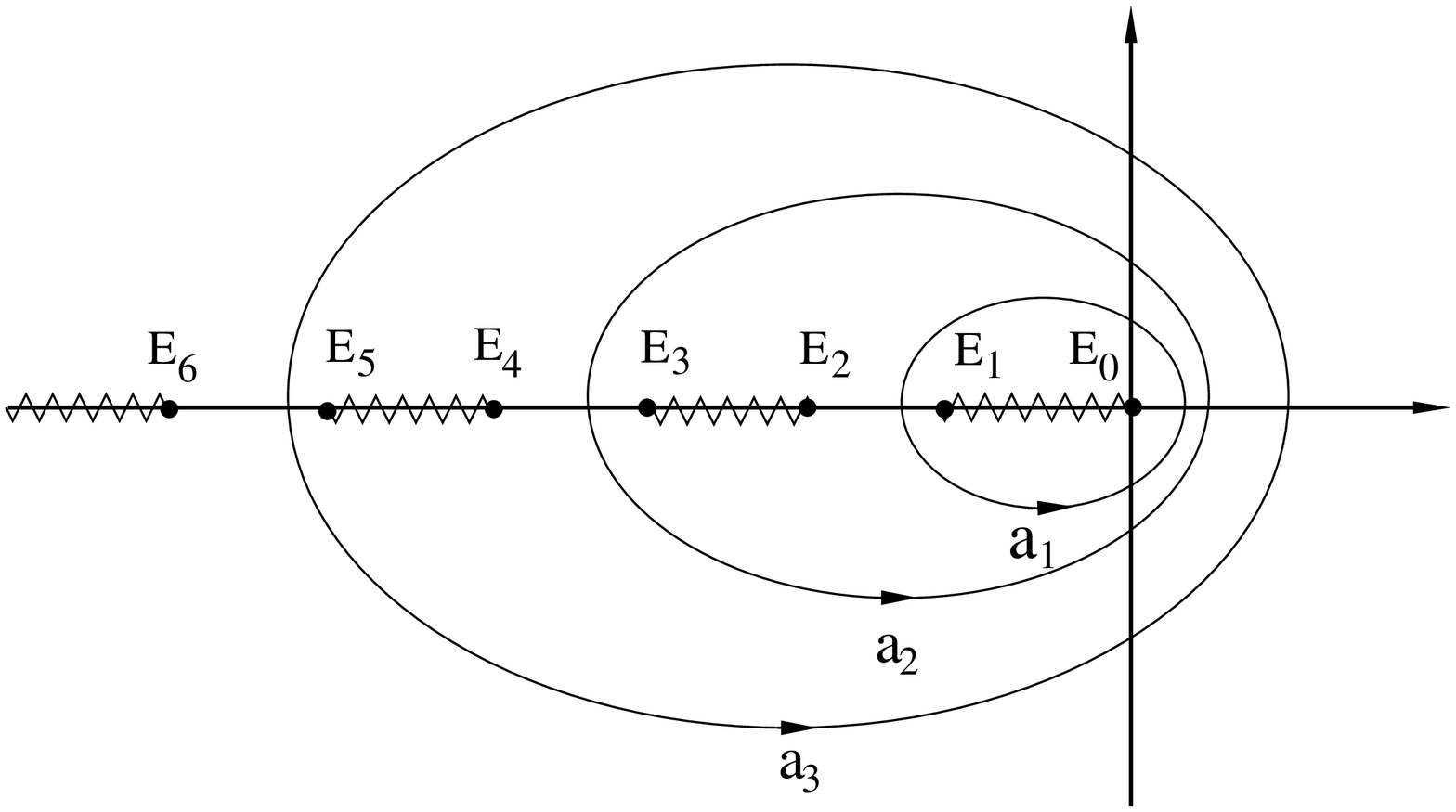}}

Fig 2.
\end{center}

We work with the following system of cuts $[E_1,E_0]$,
$[E_3,E_2]$,\ldots, $(-\infty,E_{2g}]$. Let us denote the sheet
containing the line $\lambda\in{\mathbb R}$, $\lambda>0$ $\mu>0$ by
$G_+$ and the second sheet by $G_-$ (see Fig~2).

For the generic real curves with $2m$ negative branch points
$E_1,\ldots,E_{2m}$ and $g-m$ complex adjoint pairs $E_{2j-1},E_{2j}=\bar E_{2j-1}$,
$j=m+1,\ldots,g$ we choose first $m$ $a$-cycles exactly as in the stable case.
The last $g-m$ $a$-cycles should be chosen as coverings on $\Gamma$ over the
path on the $\lambda$-plane connecting the points $E_{2j-1}$ and $E_{2j}$ for
$j>m$. All these path should not meet each other. All of them should cross
positive part of the line $\lambda>0$ in one point $\kappa_j$ (i.e. in two
points of Riemann surface) such that $\kappa_{m+1}<\kappa_{m+2}<\ldots < \kappa_g$.
This basis depends on the order of the complex conjugate pairs only (see Fig~3).

\begin{center}
\mbox{\epsfxsize=10cm \epsffile{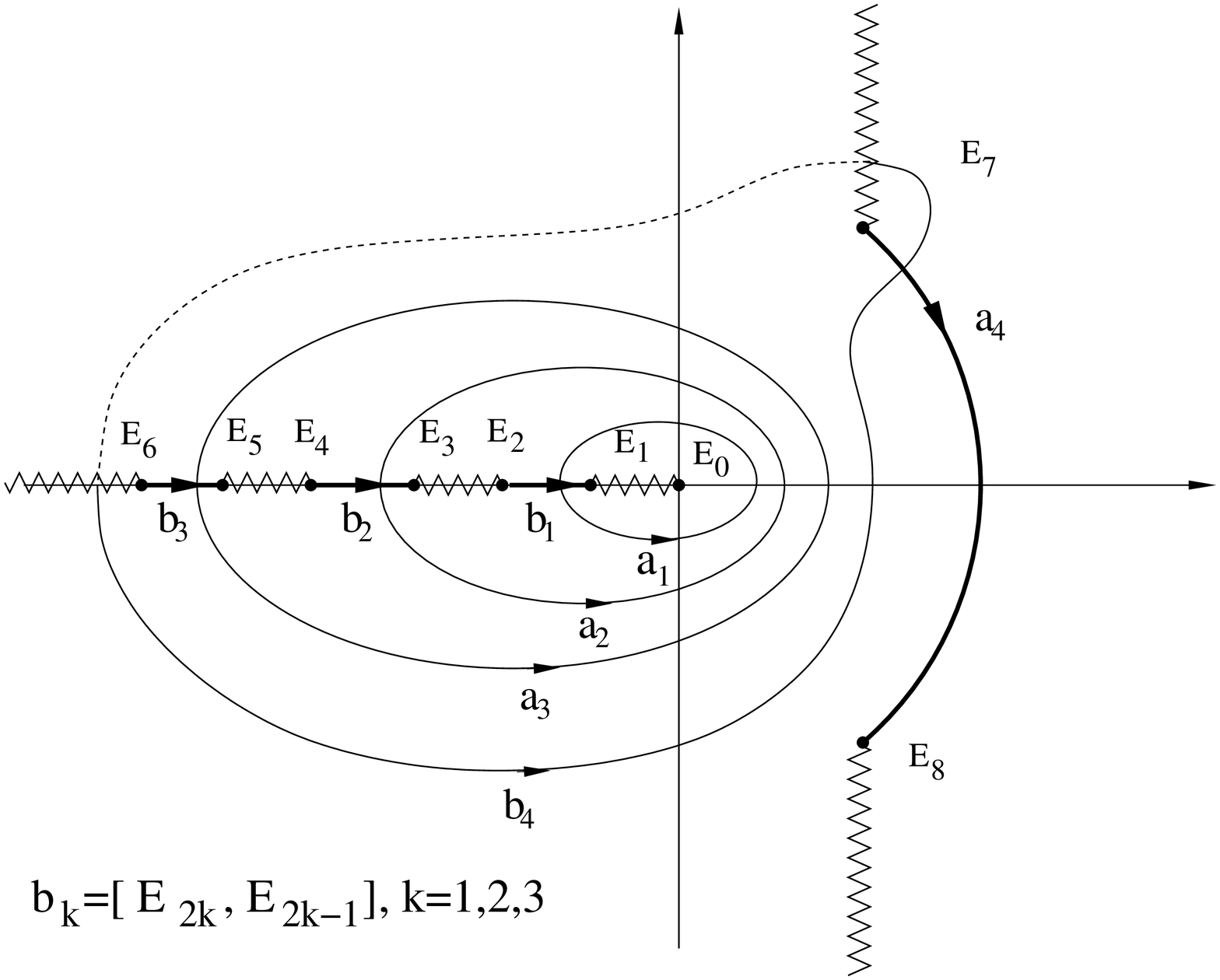}}

Fig 3.
\end{center}
We assume also that the cycles $b_1$, $b_2$, \ldots, $b_m$ are the ovals
lying over the segments $[E_2,E_1]$, $[E_4,E_3]$, $[E_{2m},E_{2m-1}]$
respectively.

For the complex conjugate pairs of branch points
$E_{2j},E_{2j-1}, j>m,$ we make the cuts connecting each of these
points with $i\infty$ or $-i\infty$ such that the cuts do not
intersect the real line. As before we denote the sheet containing
the line $\lambda\in{\mathbb R}$, $\lambda>0$ $\mu>0$ by $G_+$ and
the second sheet by $G_-$. The orientation of the cycles $b_l$,
$l=1,\dots,m$ coincides with the standard orientation of the
negative semi-line ${\mathbb R}_-$ at the sheet $G_+$; they are
opposite at the sheet $G_-$.

Consider a basic cycle ${\mathcal A}_k$ on the real component
of Jacoby torus, represented by the closed curve $B_k$ on this torus.
The image of this cycle in the surface $\Gamma$ (i.e. image of it under
the inverse Abel map) is a closed oriented curve $C_k$ (may be
consisting from several connected components) homological to the cycle
$a_k\in H_1(\Gamma,{\mathbb Z})$. From Lemma~\ref{SG-L2} it follows,
that this curve does not touch the closed segments on the real line
$[-\infty,E_{2m}],\ldots,[E_3,E_2], [E_1,0]$, and if it crosses the
negative semi-line at the inteval $[E_{2l},E_{2l-1}]$, then it happens
in the sheet $G_+$ if $(-1)^{l-1}s_l>0$ and in the sheet
$G_-$ if $(-1)^{l-1}s_l<0$. From Formula~\ref{SG-1.6} it follows, that
the topological charge along the cycle ${\mathcal A}_k$ is given by
\begin{equation}
\label{SG-2.6}
n_k=\tilde C_k \circ {\mathbb R}_-,
\end{equation}
where $\tilde C_k$ is the projection of $C_k$ to the
$\lambda$-plane, and $\circ$ denotes the
intersection index of  curves on the $\lambda$-plane. Taking into
account that any intercetion with the negative semi-line is the
intersection with one of the $b$-cycles we obtain:

\begin{lemma} The topological charge along the cycle ${\mathcal A}_k$
on the real torus in the Jacobian variety is equal to the
intersection index of the corresponding curve $C_k$ with the {\bf
Counting Cycle of the real component} on the Riemann surface
$\Gamma$
\begin{equation}
\label{SG-2.7} n_k=C_k\circ \left(\sum\limits_{l=1}^m (-1)^{l-1} s_l
b_l\right).
\end{equation}
\end{lemma}

Let us mention that the cycle $C_k$ is homotopic to $a_k$ on
$\Gamma$. Therefore $C_k\circ b_l=\delta_{kl}$ and
\begin{equation}
\label{SG-2.8}
\left\{
\begin{array}{ll}
n_k=(-1)^{k-1} s_k & \mbox{for} \ \ k\le m \\
n_k=0 & \mbox{for} \ \ k> m.
\end{array}
\right.
\end{equation}
Combining (\ref{SG-2.4}) with (\ref{SG-2.8}), we obtain our main result:

\begin{theorem}
For any real solution the density of topological charge
along the variable $x$ is given by the formula
\begin{equation}
\label{SG-2.9}
\bar n=\frac12 \ \sum\limits_{k=1}^{m} (-1)^{k-1} s_k U^k,
\end{equation}
where the components $U^k$ of the vector $\vec U$ are defined by (\ref{SG-2.2}).
\end{theorem}

\section{Averaging of the local conservation laws and topological types}
\label{section-averaging}

For a generic functional
\begin{equation}
\label{SG-3.1}
F[u]=\int f(e^{iu},e^{-iu},u_x,u_{xx},\ldots,u_t,u_{tx},u_{txx},\ldots) dx
\end{equation}
the result of averaging over a real componet of the Jacoby torus
depends on the topolgical type. For example by
avearging of $u_x$ we get the topological charge. But if we average the
Sine-Gordon Hamiltonian
\begin{equation}
\label{SG-3.2}
H[u]=\int \left[\frac{u_t^2}{2}+ \frac{u_x^2}{2}+(1-cos(u))   \right] dx
\end{equation}
the result is the same for all components. Therefore it is natural to
formulate the following:

{\bf Problem.} How to characterize the functional such, that the averaging
does not depend on the topological type?

A complete classification seems to be a complicated problem. But it is
rather easy to formulate a natural sufficient condition.

If $f(e^{iu},e^{-iu},u_x,u_{xx},\ldots,u_t,u_{tx},u_{txx},\ldots)$ is a
polynomial of its arguments then it can be written as a symmetric rational
function of the divisor coordinates $f=\tilde
f(\lambda_1(x,t),\ldots,\lambda_g(x,t),\mu_1(x,t),\ldots,\mu_g(x,t))$.
\begin{lemma}
\label{SG-L3.1}
 Consider the following form on the
 $\underbrace{\Gamma\times\Gamma\times\ldots\Gamma}_{g\ \ \mathrm
 {times}}$
 \begin{equation}
  \label{SG-3.3}
 W_f=\tilde f(\lambda_1,\ldots\lambda_g,\mu_1,\ldots,\mu_g)
  \prod\limits_{k<l}(\lambda_k-\lambda_l)\prod\limits_{k=1}^g\frac{d\lambda_k}{\mu_k}
 \end{equation}
 If we fix all divisor points except $\gamma_1$ we obtain a meromorphic
 differntial on $\Gamma$ with poles only at $0$ and $\infty$. Assume
 that the residues of this differential are equal to $0$ identically in
 $\gamma_2$,\ldots $\gamma_g$. Then for a given spectral curve the
 averaging does not depend on the topological type
\end{lemma}
{\bf Proof.} Following \cite{EFMM} we can calculate the averaging of F[u] using
the formula:
\begin{equation}
\label{SG-3.4}
\overline{F[u]}=C\oint\limits_{a_1}\ldots\oint\limits_{a_g}W_f
\end{equation}
where $C$ is the normalization constant, and {\bf all cycles $a_k$ have
the proper topological type.} The last condition was not discussed in
\cite{EFMM}. The deformations of the integration cycles do not affect
the integral if we do not cross $0$. If the condition of
Lemma~\ref{SG-L3.1} is fulfilled, then the we can move the integration
path through $0$ and ``reach'' any topological type without changing the
integral.

As a corollary we immediately get the following statement:

\begin{theorem}
Assume that the functional $F[u]$ is invariant with respect to the
 following symmetries:
 $\sigma_k:(\lambda_k,\mu_k)\rightarrow(\lambda_k,-\mu_k)$,
 $\sigma_k:(\lambda_l,\mu_l)\rightarrow(\lambda_l,\mu_l)$ for $l\ne k$.
Then the result of averaging does not depend on the topological type.
\end{theorem}

{\bf Important example.} Consider the densities of ``higher SG hamiltonians''
defined as odd expansion coefficients of the function
\begin{equation}
\label{SG-3.6}
\frac{\Psi_x(\lambda,x,t)}{\Psi(\lambda,x,t)}
\end{equation}
at the points $0$ and $\infty$.
Dircet calculation shows, that
\begin{equation}
\label{SG-3.7}
\frac{\Psi_x}{\Psi}=\frac14 \left[\frac{\Psi_\xi}{\Psi}+ \frac{\Psi_\eta}{\Psi}\right]
\end{equation}
where
\begin{equation}
\label{SG-3.8}
\frac{\Psi_\xi}{\Psi}=i\frac{\mu+Q^\xi(\lambda)}
{(\lambda-\lambda_1(x,t))\ldots(\lambda-\lambda_g(x,t))},
\end{equation}
\begin{equation}
\label{SG-3.9}
\frac{\Psi_\eta}{\Psi}=i\frac{[\mu+\lambda Q^\eta(\lambda)]
[-\lambda_1(x,t),\ldots(-\lambda_g(x,t))] }
{\lambda(\lambda-\lambda_1(x,t))\ldots(\lambda-\lambda_g(x,t))\sqrt{E_1\ldots
E_{2g}}},
\end{equation}
$Q^\xi(\lambda)$, $Q^\eta(\lambda)$ are polynomials of degree $g-1$
determined by the following conditions:
\begin{equation}
\label{SG-3.10}
Q^\xi(\lambda_k(x,t))=\mu_k(x,t), \ \
Q^\eta(\lambda_k(x,t))=\frac{\mu_k(x,t)}{\lambda_k(x,t)}.
\end{equation}
We see that the odd part of $\Psi_x \Psi^{1}$ depends on
$\lambda_k(x,t)$ but not $\mu_k(x,t)$. Therefore the averaging of the
``higher SG hamiltonians'' does not depend on the topological type.

\bibliographystyle{plain}

\end{document}